\documentclass{article}
\usepackage{spconf,amsmath,graphicx}

\usepackage{amsfonts}
\usepackage{multirow,array,enumerate,color,graphicx,fancybox,pifont,epsf,epsfig,subfigure,amsmath,amssymb,psfrag}
\usepackage{algorithm}
\usepackage{algorithmic}
\usepackage{cite}
\newcounter{MYtempeqncnt}

\newtheorem{theorem}{\textbf{Theorem}}

\newtheorem{proposition}[theorem]{\textbf{Proposition}}

\newcommand{\secref}[1]{Section~\ref{#1}}

\newcommand{\figref}[1]{Figure~\ref{#1}}

\newcommand{\theoref}[1]{Theorem~\ref{#1}}

\newcommand{\proref}[1]{Proposition~\ref{#1}}

\title{Distributed Quantization for Compressed Sensing}
\name{Amirpasha Shirazinia, Saikat Chatterjee, Mikael Skoglund}
\address{Communication Theory Department, ACCESS Linnaeus Centre, KTH Royal Institute of Technology}
\begin{document}
\ninept

\maketitle

\begin{abstract}
\vspace{-0.15cm}
We study distributed coding of compressed sensing (CS) measurements using vector quantizer (VQ). We develop a distributed framework for realizing optimized quantizer that enables encoding CS measurements of correlated sparse sources followed by joint decoding at a fusion center. The optimality of VQ encoder-decoder pairs is addressed by minimizing the sum of mean-square errors between the sparse sources and their reconstruction vectors at the fusion center. We derive a lower-bound on the end-to-end performance of the studied distributed system, and propose a practical encoder-decoder design through an iterative algorithm.
\end{abstract}
\vspace{-0.1cm}
\vspace{-0.2cm}
%%%%%%%%%%%%%%%%%%%%%%%%%%%%%%%%%%%%%%%%%%%%%%%%%%%%%%%%%%%%%%%%%%%%%%%%%%%%%%%%%%%%%%%%%%%%%%%%
\section{Introduction} \label{sec:intro}
%%%%%%%%%%%%%%%%%%%%%%%%%%%%%%%%%%%%%%%%%%%%%%%%%%%%%%%%%%%%%%%%%%%%%%%%%%%%%%%%%%%%%%%%%%%%%%%%
\vspace{-0.2cm}
With the rapid growth of sensor networks and their popularity to accomplish autonomous tasks such as sensing and computation, distributed coding of correlated sources has attracted much attention. In short, distributed coding considers separately encoding of correlated sources (from different terminals), and decoding the coded symbols jointly at a fusion center (FC). Distributed lossless and lossy source coding are well-developed fields and various theoretical results on this topic have been established \cite{73:Slepian,99:Zamir,04:Zixiang,05:Oohama,08:Wagner,09:Wernersson,13:Sun}. The characteristics of sensor networks motivate the development of new techniques and algorithms which need to be aware of the systems' limited resources, computational complexity and power consumption. In this regard, compressed sensing (CS) \cite{08:Candes} can be considered as an emerging tool for signal compression and acquisition that significantly reduces costs due to sampling and reconstruction, leading to lower power consumption and computations. By exploiting signal's sparsity in a known domain, CS allows the signal to be retrieved from relatively few measurements through a non-linear procedure.

The problem of source coding (through quantization) for CS measurements in a point-to-point setup has gained significant attention recently, and a wide range of interesting problems regarding the design and analysis of CS reconstruction and quantization algorithms have been formulated, see, e.g., \cite{06:Candes2,10:Sinan,10:Zymnis,11:Dai,11:Jacques,12:Yan,12:Kamilov,09:Sun,12:Pasha1,12:Boufounos,11:Kamilov,13:Pasha_journal,08:Goyal,12:Laska,12:Pasha1}. Considering communication channel errors, in \cite{13:Pasha_COVQ}, we have studied joint source-channel coding for CS through optimal design of vector quantizer (VQ).

All of these works are dedicated to point-to-point (single-terminal) quantized transmission of CS measurements through either noiseless or noisy communication channel. In a distributed fashion, Bajwa \textit{et al.} \cite{07:Bajwa} proposed a distributed joint source-channel communication architecture by transmitting uncoded (analog) random projections of sensor data to the FC. Feizi \textit{et al.} \cite{10:Feizi} have proposed a low-complexity and distributed lossless source compression scheme by providing a link between analog network coding and CS. This idea has been generalized in \cite{11:Feizi} as well. In this paper, similar to the above mentioned works, we consider a distributed framework for CS. However, an important difference in our work is that on the contrary to \cite{10:Feizi,11:Feizi,07:Bajwa}, we focus on \textit{lossy compression} (using VQ) and \textit{encoded transmission} of CS measurements followed by \textit{joint decoding} at the FC. To the best of our knowledge, the distributed design and analysis of VQ for CS measurements--which can be of great importance in sensor networks-- have not been addressed yet. 

In this work, we consider, without loss of generality, two linear CS systems where they obtain few number of measurements from two correlated sparse vectors. The low-dimensional (possibly noisy) measurements are encoded using VQ, and then transmitted to a FC for joint reconstruction of correlated sparse sources. As a performance criterion in this distributed setup, we are interested in the sum of mean square error (MSE) distortions between the sparse source vectors and their reconstruction vectors at the FC. By applying the end-to-end MSE criterion, our contributions are as following: \textit{(1) Establishing (necessary) conditions for optimality of VQ encoder-decoder pairs. (2) Deriving analytical expressions for the minimum mean square error (MMSE) estimators of correlated sparse sources from noisy CS measurement vectors; the MMSE estimation is an integral part of optimum VQ scheme in distributed CS setup. (3) Providing a theoretical lower-bound on the MSE performance. (4) Developing a practical VQ encoder-decoder design algorithm through an iterative algorithm.}

\vspace{-0.35cm}
%%%%%%%%%%%%%%%%%%%%%%%%%%%%%%%%%%%%%%%%%%%%%%%%%%%%%%%%%%%%%%%%%%%%%%%%%%%
\section{Distributed MMSE estimation for CS} \label{sec:dist bayesian}
%%%%%%%%%%%%%%%%%%%%%%%%%%%%%%%%%%%%%%%%%%%%%%%%%%%%%%%%%%%%%%%%%%%%%%%%%%%
\vspace{-0.15cm}
In this section, we introduce the CS system setup, and study the MMSE estimation of correlated sparse sources from noisy CS measurements in a distributed framework. We consider an exact $K$-sparse (in a known basis) vector $\mathbf{\Theta} \!\in\! \mathbb{R}^N$ comprised of $K$ non-zero coefficients ($K \! \ll \! N$). We define the (random) support set of the vector $\mathbf{\Theta} \! \triangleq  \! [\Theta_{1},\ldots,\Theta_N]^\top$ as $\mathcal{S} \! \triangleq \! \{n \in \{1,2,\ldots,N\}: \Theta_n \!\neq 0 \!\}$ with $|\mathcal{S}| \!=\! \|\mathbf{\Theta}\|_0 \!=\! K$. Here, $\|\cdot\|_0$ and $|\cdot|$ denote the $\ell_0$ norm and cardinality, respectively. The non-zero coefficients of $\mathbf{\Theta}$ are drawn according to $\Theta_n \stackrel{\text{iid}}{\! \sim}\! \mathcal{N}(0,\sigma_{\theta}^2)$, $n \!\in\! \mathcal{S}$. We assume a distributed CS system with two correlated sources $\mathbf{X}_1 , \mathbf{X}_2 \!\in\! \mathbb{R}^N$ in which the correlation is established according to
\vspace{-0.15cm}
\begin{equation} \label{eq:correlation}
    \mathbf{X}_l = \mathbf{\Theta} + \mathbf{Z}_l, \hspace{0.2cm} l \in \{1,2\},
    \vspace{-0.15cm}
\end{equation}
where $\mathbf{Z}_l \! \triangleq \! [Z_{1,l}, \!\ldots\! ,Z_{N,l}]^\top \!\in \!\mathbb{R}^N$ is an exact $K$-sparse vector with a common support set $\mathcal{S}$ whose non-zero components are drawn as $Z_{n,l} \stackrel{\text{iid}}{\sim}  \mathcal{N}(0, \sigma_{z_l}^2)$, $n \in \mathcal{S}$; thus $\|\mathbf{Z}_l\|_0 \!=\! K$, $l \!\in \!\{1,2\}$. We  assume that $\mathbf{Z}_1$ and $\mathbf{Z}_2$ are uncorrelated with each other and with the common signal $\mathbf{\Theta}$. This joint sparsity model (also known as JSM-2) for distributed CS was first introduced in \cite{09:Baron}. 

Here, we stress that although the extension to arbitrary number of correlated sources is straightforward, the assumption of a distributed system with two sources simplifies the presentation throughout the paper.
We assume, without loss of generality, that $\sigma_{z_1}^2 = \sigma_{z_2}^2 \triangleq \sigma_z^2$ and $\sigma_\theta^2 + \sigma_z^2 = 1$. To measure the amount of correlation between sources, we define the \textit{correlation ratio} as
\vspace{-0.1cm}
\begin{equation} \label{eq:corr ratio}
    \rho \triangleq \sigma_\theta^2 / \sigma_z^2.
        \vspace{-0.1cm}
\end{equation}
Hence, $\sigma_\theta^2 = \frac{\rho}{1+\rho}$ and $\sigma_z^2 = \frac{1}{1+\rho}$. Also, note that $\rho \! \rightarrow \! \infty$ implies that the sources are highly correlated, whereas $\rho \! \rightarrow \! 0$ means that they are highly uncorrelated. The remote sparse sources $\mathbf{X}_1$ and $\mathbf{X}_2$ are measured by CS-based sensors, leading to measurement vectors $\mathbf{Y}_1 \in \mathbb{R}^{M_1}$ and $\mathbf{Y}_2 \in \mathbb{R}^{M_2}$ described as
    \vspace{-0.1cm}
\begin{equation} \label{eq:meas eq1}
    \mathbf{Y}_l = \mathbf{\Phi}_l \mathbf{X}_l + \mathbf{W}_l, \hspace{0.2cm} l \in \{1,2\}, \hspace{0.1cm} \|\mathbf{X}_l\|_0 = K,
        \vspace{-0.1cm}
\end{equation}
where $\mathbf{\Phi}_l \in  \mathbb{R}^{M_l \times N}$ is a fixed sensing matrix of the $l^{th}$ sensor, and $\mathbf{W}_l \in \mathbb{R}^{M_l}$ is an additive measurement noise vector independent of other sources. Without loss of generality, we  assume that $M_1 \!= M_2 \! \triangleq \! M$, and in the sprit of CS, we consider that $K \! < \! M  \!< \! N$.

In order to reconstruct an unknown sparse source from noisy under-sampled measurement vectors in a distributed fashion, we define CS reconstruction distortion as \vspace{-0.15cm}
\begin{equation} \label{eq:CS distotion}
    D_{cs} \triangleq \frac{1}{2K} \sum_{l=1}^2 \mathbb{E}[\|\mathbf{X}_l - \widetilde{\mathbf{X}}_l\|_2^2], \vspace{-0.15cm}
\end{equation}
where $\widetilde{\mathbf{X}}_l \in \mathbb{R}^N$ ($l \in \{1,2\}$) is an estimation vector of the sparse source $\mathbf{X}_l$ from noisy CS measurements $\mathbf{Y}_1$ and $\mathbf{Y}_2$. Here, $\|\cdot\|_2$ denotes $\ell_2$ norm. In this work, we focus on the Bayesian minimum mean square error (MMSE) framework for distributed reconstruction since we are interested in the MSE performance of random sparse vectors with known priors. Later in \secref{sec:Design}, it will be shown that the MMSE estimator is central in developing distributed quantizer design procedure which is the main focus of this paper. The following proposition provides an analytical expression for the MMSE estimator. We omit the proof (given in \cite{13:Pasha_dist}) due to space limitation.
\vspace{-0.15cm}
\begin{proposition} \label{theo1}
    Consider the following assumptions: (i) the $K$-sparse sources $\mathbf{X}_1,\mathbf{X}_2$ are correlated based on the model \eqref{eq:correlation} with ratio $\rho$. (ii) The $K$ elements of the support set are uniformly drawn from all ${N \choose K}$ possibilities. (iii) The measurement noise vector is distributed as $\mathbf{W}_{\! l} \! \sim \! \mathcal{N}(\mathbf{0}, \! \sigma_{w_l}^2 \mathbf{I}_{\!M})$, $l \in \{1,2\}$, which is uncorrelated with the CS measurements and sources. Then, the MMSE estimator of $\mathbf{X}_l$ given the noisy CS measurement vector $\mathbf{y} = [\mathbf{y}_1^\top \hspace{0.15cm} \mathbf{y}_2^\top]^\top$ that minimizes $D_{cs}$ in \eqref{eq:CS distotion}, is obtained by $\widetilde{\mathbf{x}}_l^\star(\mathbf{y}) \triangleq \mathbb{E}[\mathbf{X}_l|\mathbf{y}]$ which has the following closed form expression
    \begin{equation} \label{eq:MMSE closed}
    \begin{aligned}
        &\widetilde{\mathbf{x}}^\star(\mathbf{y}) \triangleq [\widetilde{\mathbf{x}}_1^\star(\mathbf{y})^\top \hspace{0.1cm} \widetilde{\mathbf{x}}_2^\star(\mathbf{y})^\top]^\top  = \frac{\sum_{\mathcal{S} \subset \mathbf{\Omega}} \beta_\mathcal{S} \cdot \widetilde{\mathbf{x}}^\star(\mathbf{y},\mathcal{S})}{\sum_{\mathcal{S} \subset \mathbf{\Omega}} \beta_\mathcal{S}}, &
    \end{aligned}
    \vspace{-0.06cm}
    \end{equation}
    where $\widetilde{\mathbf{x}}^\star(\mathbf{y},\mathcal{S}) \triangleq \mathbb{E}[\mathbf{X}|\mathbf{y},\mathcal{S}]$ and within its support
    \begin{equation} \label{eq:dist mmse supp}
        \widetilde{\mathbf{x}}^\star(\mathbf{y},\mathcal{S}) =  \left[\begin{array}{c  c  c}
        \mathbf{I}_K & \mathbf{I}_K & \mathbf{0}_K  \\
        \mathbf{I}_K & \mathbf{0}_K & \mathbf{I}_K\\
      \end{array}\right] \mathbf{C}^\top \mathbf{D}^{-1} \mathbf{y},
    \end{equation}
    and otherwise zeros. Further,
\begin{equation*}
\begin{aligned}
\beta_\mathbf{s} &= e^{\frac{1}{2} \left(\mathbf{y}^{\! \top} (\mathbf{N}^{-\!1} \mathbf{F}^{\! \top} (\mathbf{E}^{-\!1} \!+ \mathbf{F}^{\!\top } \mathbf{N}^{-\!1}\mathbf{F})^{\!-1} \mathbf{F} \mathbf{N}^{-\!1})\mathbf{y} - \ln \det (\mathbf{E}^{-\!1} \!+ \mathbf{F}^{\! \top} \mathbf{N}^{-\!1}\mathbf{F})  \right)} & \\
    \mathbf{C} &=
  \left[\begin{array}{c  c c}
    \frac{\rho}{1 + \rho} \mathbf{\Phi}_{1,\mathcal{S}} & \frac{1}{1 + \rho} \mathbf{\Phi}_{1,\mathcal{S}} & \mathbf{0}_{M \times K} \\
    \frac{\rho}{1 + \rho} \mathbf{\Phi}_{2,\mathcal{S}} & \mathbf{0}_{M \times K} & \frac{1}{1 + \rho} \mathbf{\Phi}_{2,\mathcal{S}}  \\
  \end{array}\right],&  \\
  \mathbf{D} &= \left[\begin{array}{c c}
    \mathbf{\Phi}_{1,\mathcal{S}} \mathbf{\Phi}_{1,\mathcal{S}}^\top + \sigma_{w_1}^2 \mathbf{I}_M & \frac{\rho}{1 + \rho} \mathbf{\Phi}_{1,\mathcal{S}} \mathbf{\Phi}_{2,\mathcal{S}}^\top \\
    \frac{\rho}{1 + \rho} \mathbf{\Phi}_{2,\mathcal{S}} \mathbf{\Phi}_{1,\mathcal{S}}^\top &  \mathbf{\Phi}_{2,\mathcal{S}} \mathbf{\Phi}_{2,\mathcal{S}}^\top + \sigma_{w_2}^2 \mathbf{I}_M \\
  \end{array}\right],&  \\
  \mathbf{N} &= \left[\begin{array}{c c}
    \sigma_{w_1}^2 \mathbf{I}_M & \mathbf{0}_M \\
    \mathbf{0}_M & \sigma_{w_2}^2 \mathbf{I}_M  \\
    \end{array}\right],&  \\
  \mathbf{E} &= \left[\begin{array}{c c c}
    \frac{\rho}{1 + \rho} \mathbf{I}_K & \mathbf{0}_K & \mathbf{0}_K \\
    \mathbf{0}_K & \frac{1}{1 + \rho} \mathbf{I}_K & \mathbf{0}_K  \\
    \mathbf{0}_K & \mathbf{0}_K & \frac{1}{1 + \rho} \mathbf{I}_K \\
    \end{array}\right],& \\
  \mathbf{F} &= \left[\begin{array}{c  c c}
     \mathbf{\Phi}_{1,\mathcal{S}} &  \mathbf{\Phi}_{1,\mathcal{S}} & \mathbf{0}_{M \times K} \\
     \mathbf{\Phi}_{2,\mathcal{S}} & \mathbf{0}_{M \times K} & \mathbf{\Phi}_{2,\mathcal{S}} \end{array}\right],&
\end{aligned}
\end{equation*}
and $\mathbf{\Phi}_{l,\mathcal{S}}$ ($l \in \{1,2\}$) is formed by choosing the columns of $\mathbf{\Phi}_l$ indexed by the elements of a possible support set $\mathbf{s}$. Also, $\Omega$ denotes the set of all possible supports.
\vspace{-0.2cm}
\end{proposition}

\begin{figure*}[!t]
\normalsize
\setcounter{equation}{9}
\begin{equation} \label{eq:D1}
\begin{aligned}
   D_1(\mathbf{y}_1,i_1) &\triangleq \mathbb{E}[\|\mathbf{X}_1 - \textsf{D}_1(I_1,I_2)\|_2^2 | \mathbf{y}_1,i_1]
    \stackrel{(a)}{=} \mathbb{E}[\|\mathbf{X}_1\|_2^2|\mathbf{y}_1] + \sum_{i_2=0}^{\mathfrak{R}_2-1} P(i_2|\mathbf{y}_1) \left[\|\textsf{D}_1(i_1,i_2)\|_2^2  - 2  \mathbb{E}[\mathbf{X}_1^\top | \mathbf{y}_1,i_2] \textsf{D}_1(i_1,i_2) \right]& \\
    &\stackrel{(b)}{=} \mathbb{E}[\|\mathbf{X}_1\|_2^2|\mathbf{y}_1] + \sum_{i_2=0}^{\mathfrak{R}_2-1}   \left[\int_{\mathbf{y}_2 \in \mathcal{R}^{i_2}} \left(\|\textsf{D}_1(i_1,i_2)\|_2^2 -2 \mathbb{E}[\mathbf{X}_1^\top|\mathbf{y}_1, \mathbf{y}_2] \textsf{D}_1(i_1,i_2) \right) p(\mathbf{y}_2|\mathbf{y}_1) d\mathbf{y}_2 \right]&
\end{aligned}
\vspace{-0.2cm}
\end{equation}
\setcounter{equation}{\value{MYtempeqncnt}}
\vspace{-0.4cm}
\hrulefill
\end{figure*}

\begin{figure*}[!t]
\normalsize
\setcounter{equation}{10}
\begin{equation} \label{eq:final enc1}
\begin{aligned}
    i_1^\star &= \text{arg }\underset{i_1 \in \mathcal{I}_1}{\text{min }} \left\{ \sum_{i_2=0}^{\mathfrak{R}_2-1}  \int_{\mathcal{R}^{i_2}} \left[\|\textsf{D}(i_1,i_2)\|_2^2 - 2 \widetilde{\mathbf{x}}^\star(\mathbf{y}_1,\mathbf{y}_2)^\top \textsf{D}(i_1,i_2) \right] p(\mathbf{y}_2|\mathbf{y}_1) d\mathbf{y}_2 \right\},
\end{aligned}
\end{equation}
\setcounter{equation}{\value{MYtempeqncnt}}
\hrulefill
\end{figure*}
\setcounter{equation}{7}

Note that the MSE of the MMSE estimator \eqref{eq:MMSE closed} can be empirically computed using Monte-Carlo simulations. It should be also mentioned that as $N$ increases, the implementation of the MMSE estimator may not feasible since the summation in \eqref{eq:MMSE closed} is taken over all possible supports. Here, we do not focus on approximation methods for the MMSE estimator because it is beyond the scope of the paper. Interested readers are referred to \cite{09:Elad} for the clues regarding the implementation of approximated MMSE estimator.

In the next section, we consider the main scenario of this work which is distributed quantization of CS measurements.
%%%%%%%%%%%%%%%%%%%%%%%%%%%%%%%%%%%%%%%%%%%%%%%%%%%%%%%%%%%%%%%%%%%%
\vspace{-0.3cm}
\section{System Description} \label{sec:descrpn}
\vspace{-0.2cm}
%%%%%%%%%%%%%%%%%%%%%%%%%%%%%%%%%%%%%%%%%%%%%%%%%%%%%%%%%%%%%%%%%%%%
In this section, we give an account for the basic assumptions and models made about the studied distributed coding system.

We consider that VQ encoders at terminals 1 and 2 (without any collaboration) be fed by the noisy CS measurements $\mathbf{Y}_1$ and $\mathbf{Y}_2$ (under model \eqref{eq:meas eq1}), respectively. The encoder mapping $\textsf{E}_l$ ($l \in \{1,2\}$) encodes $\mathbf{Y}_l$ to a transmission index $i_l$, i.e., $\textsf{E}_l: \mathbb{R}^M \rightarrow \mathcal{I}_l$ where $i_l \in \mathcal{I}_l$, and $\mathcal{I}_l$ is a finite index set defined as $\mathcal{I}_l \triangleq \{0,1,\! \ldots \!,2^{R_l}-1\}$ with $|\mathcal{I}_l| \! \triangleq \!\mathfrak{R}_l\! =\! 2^{R_l}$. Here, $R_l$ is the assigned quantization rate for each encoder in bits/vector. We fix the total quantization rate at $R_1 \!+\! R_2 \triangleq R$ bits/vector. Denoting the encoded index by $I_l$, the encoders are specified by the regions $\{\mathcal{R}_{i_l}\}_{i_l=0}^{\mathfrak{R}_l-1}$ such that when $\mathbf{Y}_l \! \in \! \mathcal{R}_{i_l}$, the encoder outputs $\textsf{E}_l(\mathbf{Y}_l)\! =\! i_l \in \mathcal{I}_l$.

Now, we consider the VQ decoder at a FC which uses both indexes $i_1 \in \mathcal{I}_1$ and $i_2 \in \mathcal{I}_2$ in order to make the estimate  of the sparse source vector, denoted by $\widehat{\mathbf{X}}_l \in \mathbb{R}^N$. Given the received indexes $i_1$ and $i_2$, the decoder is a mapping $\textsf{D}_l: \mathcal{I}_1 \times \mathcal{I}_2 \rightarrow \mathcal{C}_l$, where $\mathcal{C}_l$, with $|\mathcal{C}_l| = 2^{R_1 + R_2}$, is a finite discrete \textit{codebook} set containing all reproduction \textit{codevectors}. The decoder's functionality is described by a look-up table; $(I_1 = i_1, I_2 = i_2) \Rightarrow (\widehat{\mathbf{X}}_1 = \textsf{D}_1(i_1,i_2),\widehat{\mathbf{X}}_2 = \textsf{D}_2(i_1,i_2))$.

We assess the end-to-end performance of our studied system by
\begin{equation} \label{eq:MSE} \setcounter{equation}{7}
    D \triangleq \frac{1}{2K} \sum_{l=1}^2 \mathbb{E}[\|\mathbf{X}_l - \widehat{\mathbf{X}}_l \|_2^2].
\end{equation}
Note that the MSE depends on \textit{CS reconstruction distortion} and \textit{quantization error}. Our goal is to design robust VQ encoder-decoder pairs against all these kinds of error.

\vspace{-0.2cm}
%%%%%%%%%%%%%%%%%%%%%%%%%%%%%%%%%%%%%%%%%%%%%%%%%%%%%%%%%%%%%%%%%%%%
\section{Distributed Quantizer Design and Analysis} \label{sec:Design}
%%%%%%%%%%%%%%%%%%%%%%%%%%%%%%%%%%%%%%%%%%%%%%%%%%%%%%%%%%%%%%%%%%%%
\vspace{-0.2cm}
We note that the joint design of the encoder and decoder mappings, $\textsf{E}_l$ and $\textsf{D}_l$ ($l \in \{1,2\}$), is generally not an easy task. Therefore, we optimize each mapping (with respect to minimizing the MSE in \eqref{eq:MSE}) by fixing the other mappings. The resulting mappings fulfil necessary conditions for optimality, and can be implemented in practice. Since the system is symmetric, we only show the optimization method for the first encoder and then the decoder.

Keeping the mappings $\textsf{E}_2$, $\textsf{D}_1$ and $\textsf{D}_2$ fixed, we have % Let us first define
\begin{equation} \label{eq:MSE der}
\begin{aligned}
    D &= \frac{1}{2K}\sum_{i_1=0}^{\mathfrak{R}_1-1} \int_{\mathbf{y}_1 \in \mathcal{R}_{i_1}} \{ \overbrace{\mathbb{E}[\|\mathbf{X}_1 - \textsf{D}_1(I_1,I_2)\|_2^2 | \mathbf{y}_1,i_1]}^{\triangleq D_1(\mathbf{y}_1,i_1)}  & \\
    &+  \underbrace{\mathbb{E}[\|\mathbf{X}_2 - \textsf{D}_2(I_1,I_2)\|_2^2 | \mathbf{y}_1,i_1]}_{\triangleq D_2(\mathbf{y}_1,i_1)}  \} p(\mathbf{y}_1) d\mathbf{y}_1,&
\end{aligned}
\end{equation}
where $p(\mathbf{y}_1)$ is the $M$-fold probability density function (pdf) of the measurement vector $\mathbf{Y}_1$. Since $p(\mathbf{y}_1)$ is a non-negative value, in order to optimize the mapping $\textsf{E}_1$ in the sense of minimizing $D$, it suffices to minimize the expression inside the braces in \eqref{eq:MSE der}. Thus, the optimized encoding index $i_1^\star$ is obtained by
\begin{equation} \label{eq:min index 1}
    i_1^\star = \text{arg }\underset{i_1 \in \mathcal{I}_1}{\text{min }} \left\{ D_1(\mathbf{y}_1,i_1) + D_2(\mathbf{y}_1,i_1)\right \}.
\end{equation}
Now, $D_1(\mathbf{y}_1,i_1)$ can be rewritten as \eqref{eq:D1} on top of the page, where $(a)$ follows by expanding the conditional expectation and the fact that $\mathbf{X}_1$ and $\textsf{D}_1(I_1,I_2)$ are independent conditioned on $\mathbf{y}_1,i_1,i_2$. Further, $(b)$ follows from marginalization of the expression inside the brackets in $(a)$ over $i_2$ and $\mathbf{y}_2$. In a same fashion, $D_2(\mathbf{y}_1,i_1)$ can be parameterized similar to \eqref{eq:D1} with the only difference that $\mathbf{X}_1$ and $\textsf{D}_1(i_1,i_2)$ are replaced with $\mathbf{X}_2$ and $\textsf{D}_2(i_1,i_2)$, respectively. Following \eqref{eq:min index 1} and \eqref{eq:D1}, the MSE-minimizing encoding index $i_1^\star$ is given by \eqref{eq:final enc1}, where $\widetilde{\mathbf{x}}^\star(\mathbf{y}_1,\mathbf{y}_2)  \triangleq \left[\widetilde{\mathbf{x}}_1^\star(\mathbf{y}_1,\mathbf{y}_2)^{\top} \hspace{0.1cm} \widetilde{\mathbf{x}}_2^\star (\mathbf{y}_1, \mathbf{y}_2)^{ \top} \right]^{ \top} $ and $\textsf{D}(i_1,i_2)  \triangleq  \left[\textsf{D}_1(i_1,i_2)^\top \hspace{0.1cm} \textsf{D}_2(i_1,i_2)^\top \right]^\top  $. Here, the codevectors $\textsf{D}_l(i_1,i_2)$, ($l \in \{1,2\}$), are given, and the vector $\widetilde{\mathbf{x}}_l^\star(\mathbf{y}_1,\mathbf{y}_2)$ denotes the MMSE estimators derived in \proref{theo1}. It should be mentioned that although the observation at terminal 2, $\mathbf{y}_2$, appears in the formulation of the optimized encoder at terminal 1, it is finally integrated out.

Assuming all encoders are fixed, it can be shown that the MSE-minimizing decoder is given by
\setcounter{equation}{11}
\begin{equation} \label{eq:final dec}
\begin{aligned}
    \textsf{D}_l^\star(i_1,i_2) \!=\! \mathbb{E}[\mathbf{X}_l | i_1,i_2]
    \!=\! \frac{\int_{\mathcal{R}^{i_1}} \! \int_{\mathcal{R}^{i_2}} \widetilde{\mathbf{x}}_l^\star (\mathbf{y}_1,\mathbf{y}_2)  p(\mathbf{y}_1,\mathbf{y}_2) d\mathbf{y}_1 d \mathbf{y}_2}{\int_{\mathcal{R}^{i_1}} \! \int_{\mathcal{R}^{i_2}} p(\mathbf{y}_1,\mathbf{y}_2) d\mathbf{y}_1 d \mathbf{y}_2},
\end{aligned}
\end{equation}
where the second equality can be shown by marginalizing of the conditional expectation over $\mathbf{Y}_1$ and $\mathbf{Y}_2$ and using the Bayes' rule.

When there is no correlation between sources ($\rho \rightarrow 0$), then it can be shown that the optimized encoder \eqref{eq:final enc1} and the optimized decoder in \eqref{eq:final dec} boil down to the optimized encoder and decoder in the
point-to-point source coding of CS measurements, cf. \cite{13:Pasha_COVQ}.

We emphasize that we do not assume any sparse structure on the reconstructed vectors at the receiving-ends in contrast to conventional $\ell_1$-norm reconstruction methods. The reason is due to the fact that we are interested in the final reconstruction MSE, where considering any kind of sparse structure might degrade the performance.

Now, we analyze the end-to-end MSE for our studied distributed system. Recall the MMSE estimation of the correlated sources $\widetilde{\mathbf{X}}_l^\star \! \triangleq \! \mathbb{E}[\mathbf{X}_l|\mathbf{Y}_1,\mathbf{Y}_2]$, $l \! \in \! \{1,2\}$, then we rewrite the end-to-end MSE as
\begin{equation} \label{eq:MSE decomp}
\begin{aligned}
    &D\!\stackrel{(a)}{=} \! \frac{1}{2K}\sum_{l=1}^2 \!\mathbb{E}[\|\mathbf{X}_l \!- \! \widetilde{\mathbf{X}}_l^\star\|_2^2] \!\!+\! \frac{1}{2K}\sum_{l=1}^2 \! \mathbb{E}[\|\widetilde{\mathbf{X}}_l^\star \!-\! \widehat{\mathbf{X}}_l\|_2^2] \! \triangleq \! D_{cs} \!+\! D_q,&
\end{aligned}
\end{equation}
where $(a)$ can be shown by the definition of the MMSE estimator $\widetilde{\mathbf{X}}_l^\star$ and by using the Markov property $\mathbf{X}_l \! \rightarrow \! (I_1,I_2)\! \rightarrow \! \widehat{\mathbf{X}}_l$, $l \in \{1,2\}$. Interestingly, \eqref{eq:MSE decomp} implies that, without loss of optimality, the end-to-end MSE, denoted by $D$, can be summed up as CS reconstruction MSE (of the MMSE estimator), denoted by $D_{cs}$, and quantization MSE, denoted by $D_{q}$. This property together with information theoretic results in \cite{08:Wagner} can be used to develop a lower-bound on $D$ provided by the following theorem. We omit the proof (given in \cite{13:Pasha_dist}) due to lack of space.

\begin{theorem} \label{theo2}
\vspace{-0.25cm}
Consider the assumptions given in \proref{theo1}. Let the total quantization rate be $R = R_1 + R_2$ bits/vector where $R_l$ is the assigned quantization rate at terminal $l \in \{1,2\}$, then the asymptotic (in quantization rate) end-to-end MSE \eqref{eq:MSE} is lower-bounded as
\begin{equation} \label{eq:total lb}
\begin{aligned}
	D \geq  \max \left\{D_q^{(lb)} , D_{cs} \right\},
\end{aligned}
\vspace{-0.2cm}
\end{equation}
where $D_{q}^{(lb)} = $
\begin{equation} \label{eq:lb sum}
	\!\sqrt{\left(1\!-\! \frac{\rho^2}{(1+\rho)^2}\right) 2^{\frac{-2\left(R -  \log_2 {N \choose k}\right)}{K}} \!+\! \frac{\rho^2}{(1+\rho)^2} 2^{\frac{-4\left(R -  \log_2 {N \choose k}\right)}{K}}}.
\end{equation}
and $D_{cs}$ is calculated by \eqref{eq:CS distotion} corresponding to the distortion due to MMSE estimation of correlated sparse sources $\mathbf{X}_1$ and $\mathbf{X}_2$ from noisy CS measurements derived in \eqref{eq:MMSE closed} of \proref{theo1}.

\vspace{-0.2cm}
\end{theorem}

When CS measurements are noisy, i.e. $\sigma_{w_1}^2, \sigma_{w_2}^2 \! \neq \! 0$, it can be verified as the quantization rate $R$ increases, the end-to-end MSE saturates to $D_{cs}$ since $D_q^{(lb)}$ decays exponentially, but $D_{cs}$ becomes constant by quantization rate. The source correlation ratio $\rho$ also plays an important role on the level of the lower-bound \eqref{eq:total lb}. By taking the first derivative of $D_q^{(lb)}$ in \eqref{eq:lb sum} with respect to $\rho$, it can be verified that the derivative is always negative which means as the source correlation $\rho$ increases, the lower-bound decreases.

\begin{figure*}[ht]
 \centering
 \subfigure[{\scriptsize CS reconstruction distortion $D_{cs}$ vs. $\rho$.}]{
   \includegraphics[width=0.66\columnwidth,height=4.5cm]{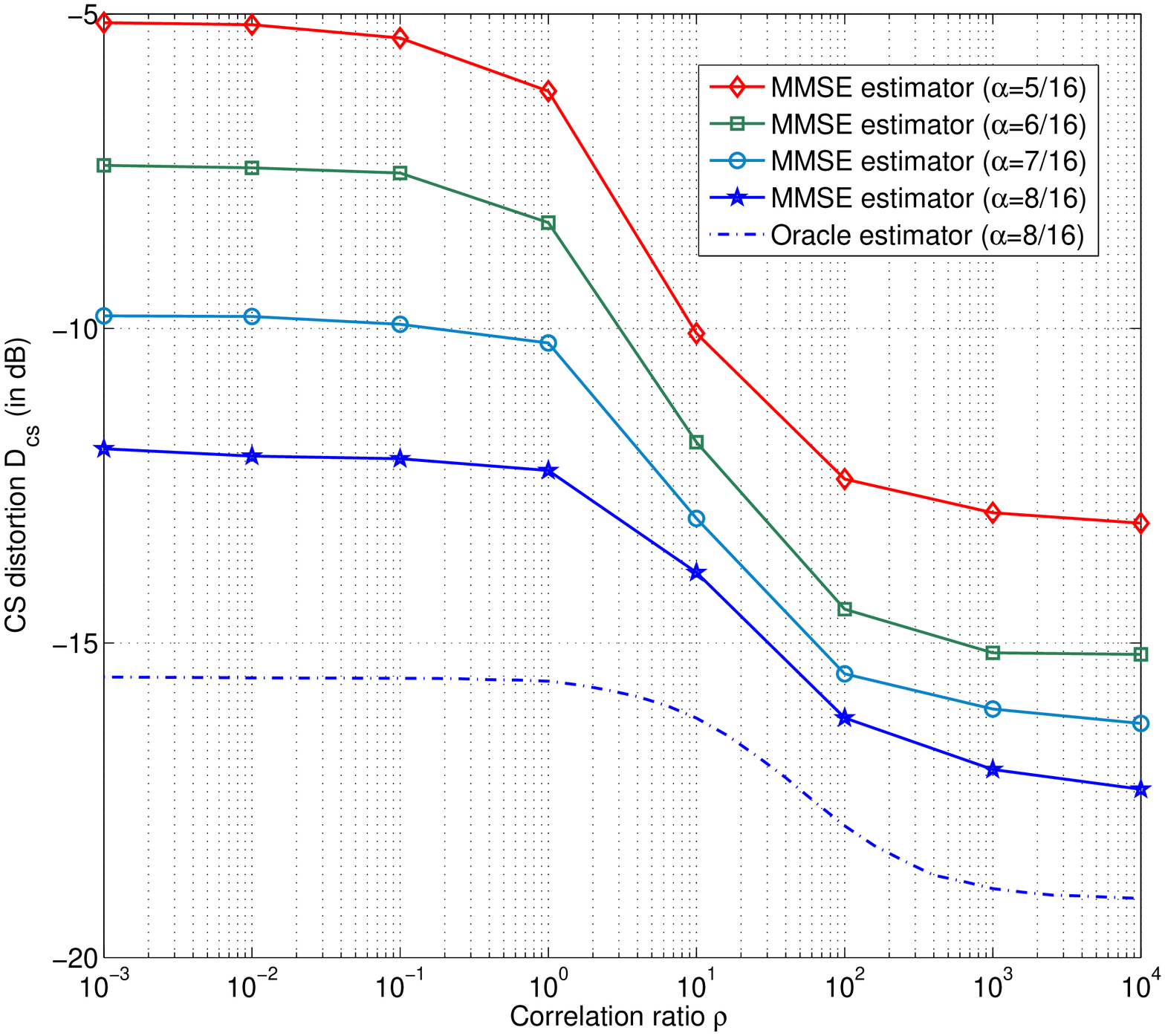}
  \label{fig:Bayes_MMSE}%\vspace{-0.4cm}
  }
 \subfigure[{\scriptsize End-to-end distortion $D$ vs. $\rho$.}]{
  \includegraphics[width=0.66\columnwidth,height=4.5cm]{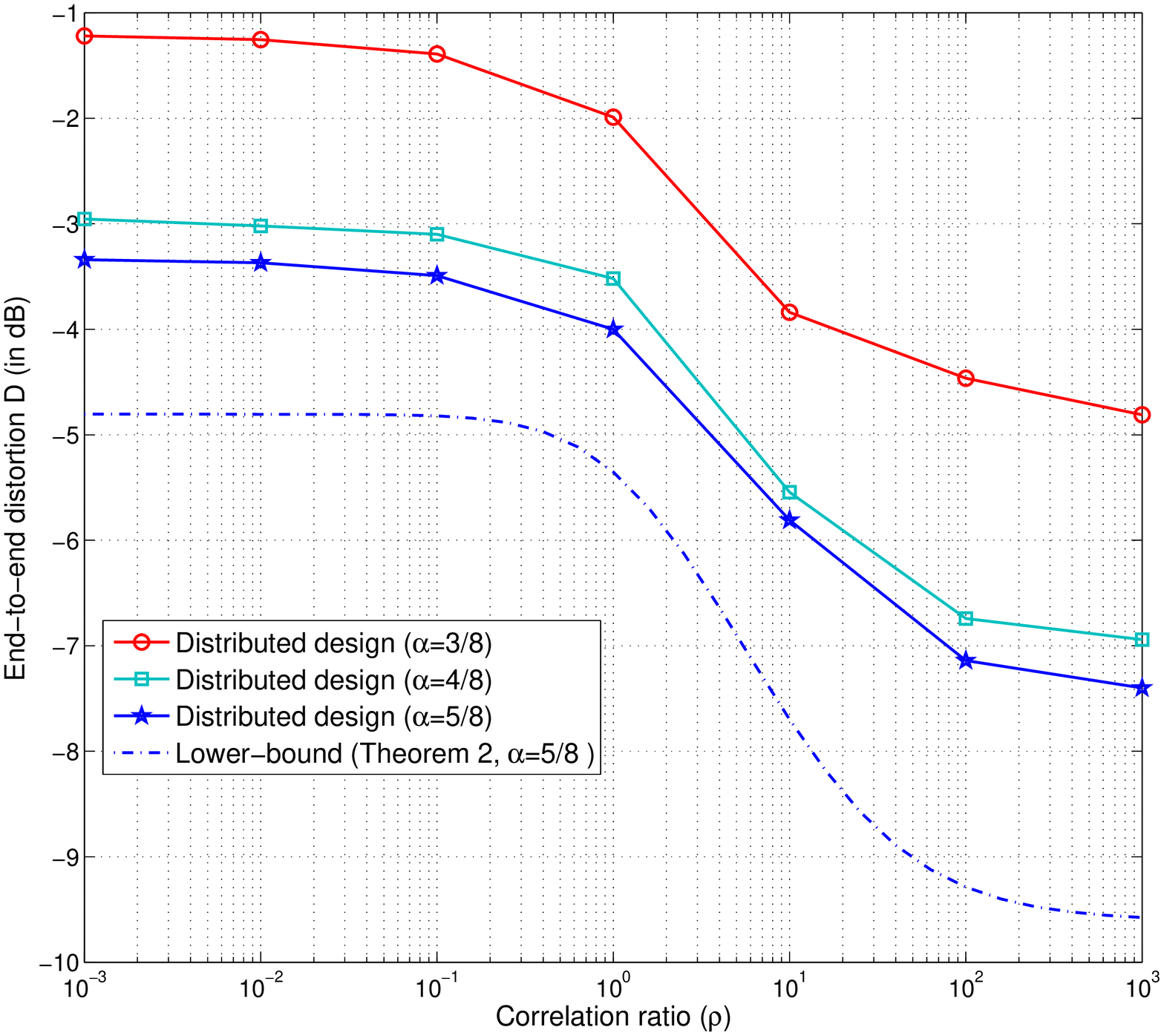}
  \label{fig:MSE_rho_8} %\vspace{-0.4cm}
  }
  \subfigure[{\scriptsize End-to-end distortion $D$ vs. quantization rate $R$.}]{
  \includegraphics[width=0.66\columnwidth,height=4.5cm]{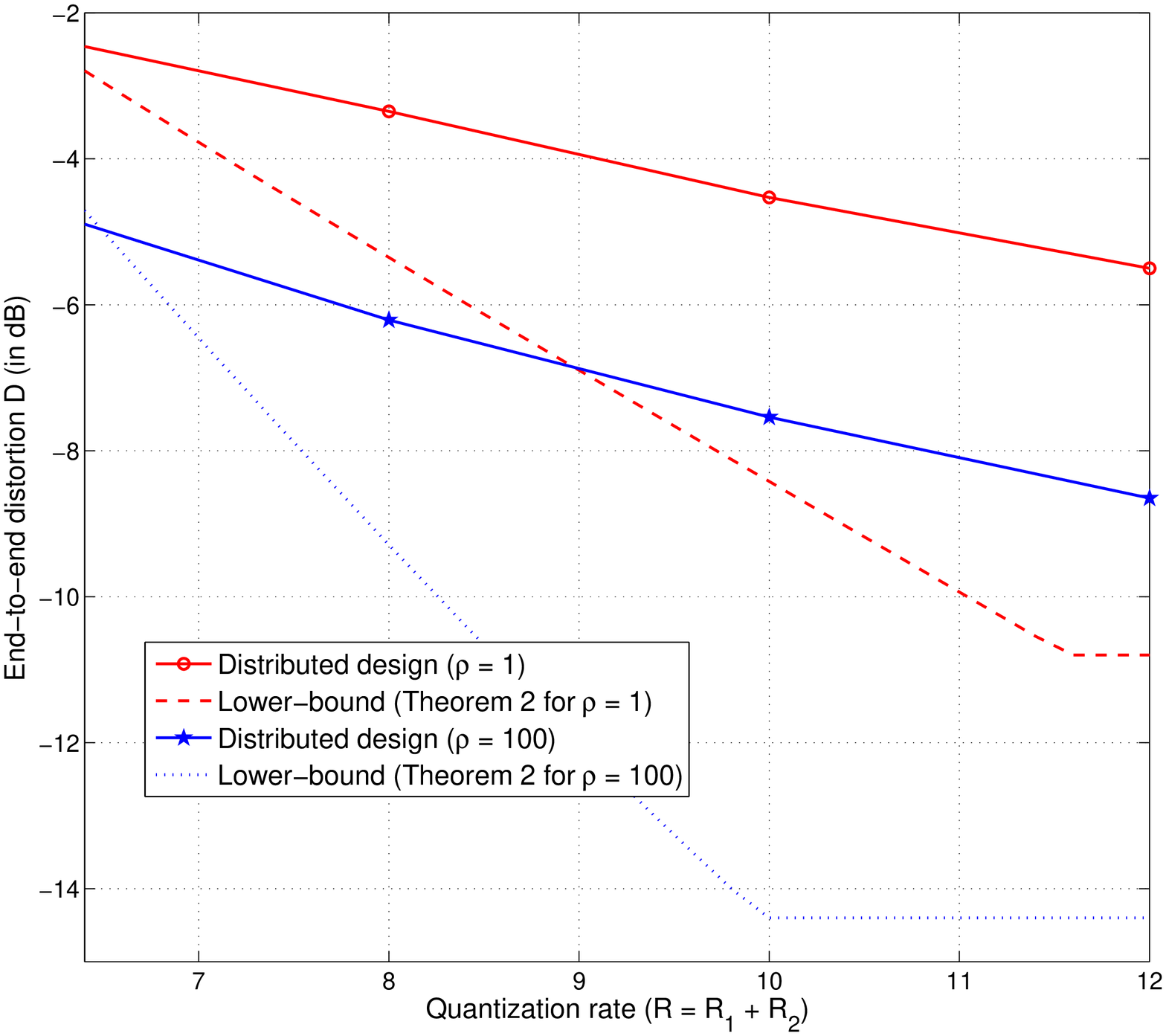}
  \label{fig:MSE_vs_R} %\vspace{-0.4cm}
  } \vspace{-0.25cm}
 \caption{ {\scriptsize Simulation results for the distributed design methods are compared with corresponding lower-bounds.}}
  \label{fig:codevec}
    \vspace{-0.45cm}
\end{figure*}

\vspace{-0.25cm}
%%%%%%%%%%%%%%%%%%%%%%%%%%%%%%%%%%%%%%%%%%%%%%%
\section{Practical Design Algorithm} \label{sec:training}
%%%%%%%%%%%%%%%%%%%%%%%%%%%%%%%%%%%%%%%%%%%%%%%
\vspace{-0.2cm}

The necessary optimal conditions in \eqref{eq:final enc1} (and its equivalence $i_2^\star$) and \eqref{eq:final dec} can be combined in an \textit{alternate-iterate} training algorithm in order to design optimized encoder-decoder pairs for distributed quantization of CS measurements which converges to locally optimum encoder regions and decoder codevectors. A natural order to optimize the mappings is: \textit{1) the first encoder, 2) the first decoder, 3) the second encoder and 4) the second decoder}.

To implement the algorithm, we make some modifications. The integral in \eqref{eq:final enc1} (for calculating the optimized encoder) cannot be solved in closed form in general. Hence, we modify the integral, and compute it numerically. The integral in \eqref{eq:final enc1} can be approximated as
\begin{equation} \label{eq:modify enc1}
\begin{aligned}
    & \|\textsf{D}(i_1,i_2)\|_2^2 P(i_2 | \widetilde{\mathbf{y}}_1) \!-\! 2 P(i_2 | \widetilde{\mathbf{y}}_1) \mathbb{E} [\mathbf{X}^\top | \widetilde{\mathbf{y}}_1 , i_2] , \textsf{D}(i_1,i_2)&
\end{aligned}
\end{equation}
where we have only approximated $\mathbf{y}_1$ by its scalar-quantized representation, denoted by  $\widetilde{\mathbf{y}}_1$, using $r_y$-bit nearest-neighbor coding (using, e.g., LBG algorithm \cite{80:LBG}). Hence, $\mathbf{y}_1$ is discretized, and $P(i_2 | \widetilde{\mathbf{y}}_1) \triangleq \text{Pr}\{I_2 = i_2 | \widetilde{\mathbf{Y}}_1 = \widetilde{\mathbf{y}}_1\}$ indicates an element of a  transition probability matrix whose transitions can be numerically computed. In order to evaluate the conditional mean $\mathbb{E}[\mathbf{X}^\top | \widetilde{\mathbf{y}}_1 , i_2]$ in \eqref{eq:modify enc1}, we generate samples of $\mathbf{X}_1$ and $\mathbf{X}_2$, and then take average over those samples that have resulted in the quantized value $\widetilde{\mathbf{y}}_1$ and the quantization index $i_2$. Using these modifications, the encoder computational complexity grows at most like $\mathcal{O}(2^{R_1+R_2})$.

Moreover, the calculation of the codevectors $\textsf{D}_l(i_1,i_2)$ derived in \eqref{eq:final dec} requires massive integrations of highly non-linear functions. Therefore, we calculate $\mathbb{E}[\mathbf{X}_l | i_1,i_2]$ empirically by generating Monte-Carlo samples of $\mathbf{X}_l$, and then take average over those samples which have led to the quantized indexes $i_1$ and $i_2$.

\vspace{-0.35cm}
%%%%%%%%%%%%%%%%%%%%%%%%%%%%%%%%%%%%%%%%%%%%%%%%%%%%%%%%%%%%%%%%%%%%%%%%%	
\section{Experiments} \label{sec:numerical}
%%%%%%%%%%%%%%%%%%%%%%%%%%%%%%%%%%%%%%%%%%%%%%%%%%%%%%%%%%%%%%%%%%%%%%%%%
\vspace{-0.25cm}

We assess the performance using CS reconstruction MSE, $D_{cs}$, and end-to-end MSE, $D$, characterized in \eqref{eq:CS distotion} and \eqref{eq:MSE}, respectively. The correlated sources with correlation ratio $\rho$ (defined in \eqref{eq:corr ratio}) are randomly generated according to the models described in \secref{sec:dist bayesian}. The sensing matrices $\mathbf{\Phi}_1$ and $\mathbf{\Phi}_2$ are produced by choosing the first (indexed from the first row downwards) and the last (indexed from the last row upwards) $M$ rows of a $N \times N$ discrete cosine transform (DCT) matrix. Then, the columns of the resulting matrices are normalized to unit-norm. To measure the level of under-sampling, we define the \textit{measurement rate} $0 \!< \!\alpha \!\leq\! 1$ as $\alpha \! \triangleq \! M/N$. We define signal-to-measurement noise ratio (SMNR) at terminal $l \!\in \! \{1,2\}$ as $\text{SMNR}_l \!\triangleq\! \mathbb{E}[\|\mathbf{X}_l\|_2^2]/\mathbb{E}[\|\mathbf{W}_l\|_2^2] \!=\! K/(M\sigma_{w_l}^2)$. All simulations are performed by generating $3 \!\times\! 10^5$ realizations of the source vectors.

In our first experiment, we study the impact of source correlation $\rho$ and measurement rate $\alpha$ on the CS reconstruction MSE, $D_{cs}$. We consider ($N=16$, $K=2$, $\text{SMNR}_l = 10$ dB), and empirically compute $D_{cs}$ for the MMSE estimator derived in \eqref{eq:MMSE closed} of \proref{theo1}. The results are illustrated in \figref{fig:Bayes_MMSE} as a function of $\rho$ for measurement rates $\alpha=5/16, \ldots, 8/16$. The empirical \textit{oracle estimator} lower-bound corresponding to the measurement rate $\alpha = 8/16$ is also demonstrated. Note that the \textit{ideal} oracle estimator is calculated from  \eqref{eq:dist mmse supp} given the \textit{a priori} known support for each source realization. From \figref{fig:Bayes_MMSE}, we observe that increasing number of CS measurements improve the performance which is expected since the sources are estimated from more amount of information. Another point is that $D_{cs}$ varies significantly by changing the correlation ratio $\rho$ which is also reflected from the oracle lower-bound. This is due to the fact that at low correlation, the measurement vectors become uncorrelated, therefore there is no gain obtained by, e.g., estimation of $\mathbf{X}_1$ from observations at the second terminal, i.e., $\mathbf{y}_2$. On the other hand, when the sources are highly correlated, the estimation procedure tends to estimating a single source $\mathbf{\Theta}$ from $2M$ observations, i.e., $\mathbf{y}_1$ and $\mathbf{y}_2$.

In our second experiment, we demonstrate the effect of $\rho$ and $\alpha$ on the end-to-end performance $D$. We use ($N = 8, K = 2, R = R_1 + R_2 = 8$ bits/vector with $R_1 = R_2$), and assume clean measurements. Further, The vectors $\mathbf{y}_1$ and $\mathbf{y}_2$ are pre-quantized using $r_y=3$ bits per measurement entry. We vary the correlation ratio from very low $\rho = 10^{-3}$ to very high values $\rho = 10^3$, and compare the simulation results with the lower-bound derived in \eqref{eq:total lb} of \theoref{theo2}. The results are shown in \figref{fig:MSE_rho_8} for measurement rates $\alpha = \frac{3}{8}, \frac{4}{8}, \frac{5}{8}$. As would be expected, at a fixed quantization rate $R$ and correlation ratio $\rho$, increasing $\alpha$ improves the performance since the sources are reconstructed from larger sets of information. Hence, the end-to-end MSE decreases, and the curves approach the lower-bound. The correlation is a useful factor to reduce the end-to-end distortion which is also reflected from the lower-bound. This behavior can be interpreted as follows. When the sources are fully correlated, $\rho \rightarrow \infty$, the two sources can be viewed as a single source, and the FC is able to reconstruct the source jointly from two sets of received quantized indexes. Hence, the performance is maximized. On the other hand, when the sources are uncorrelated, $\rho \rightarrow 0$, the FC reconstructs the sources from two sets of independent quantized indexes. Thus, there is no gain in joint decoding at the FC.

Now, we investigate how the performance varies by quantization rate. We use the simulation parameter set $(N = 8, K = 2, \alpha = 5/8, \text{SMNR}_1 = \text{SMNR}_2 = 10$) dB. In \figref{fig:MSE_vs_R}, we illustrate the end-to-end MSE of the proposed design method as a function of total quantization rate $R = R_1 + R_2$ (with $R_1 = R_2$) for two values of correlation ratios: $\rho = 1$ (low-correlated sources) and $\rho = 100$ (high-correlated sources). The simulation curves are compared with the lower-bound in \eqref{eq:total lb} corresponding to $\rho=1,100$. From \figref{fig:MSE_vs_R}, we observe that the performance improves by increasing quantization rate. Moreover, increasing correlation between sources reduces the MSE as observed from other experiments too. We also note that as quantization rate increases, all the curves converge to their respective MSE floors, specified by $D_{cs}$ that can be seen from the lower-bounds in \figref{fig:MSE_vs_R}.

\vspace{-0.45cm}	
\section{Conclusions} \label{sec:conclusion}
\vspace{-0.25cm}

We studied the design and analysis of the distributed vector quantization of CS measurements. We derived necessary conditions for optimality of encoder-decoder pairs by minimizing end-to-end MSE. We analyzed the MSE and proved that it is the sum of CS reconstruction MSE (of MMSE estimator) and quantization MSE. This result helped us to derive a lower-bound on the end-to-end MSE. Simulations revealed that correlation between sources, besides compression resources e.g. measurement and quantization rates, is an effective factor on CS reconstruction MSE as well as quantization MSE.

\vspace{-0.35cm}
\bibliographystyle{IEEEtran}
\bibliography{IEEEfull,bibliokthPasha}
\end{document}